\documentclass[prb,twocolumn,amsmath,amssymb,superscriptaddress]{revtex4-2}

\usepackage{graphicx}
\usepackage{bm}
\usepackage{color}
\usepackage{url}
\usepackage{tikz}
\usetikzlibrary{calc}
\usepackage{pgfplots}
\usepackage{amsmath,amssymb}
\usepackage{verbatim}
\usepackage{physics}
\usepackage{xcolor}
\usepackage{mwe}
\usepackage{tikz}
\usetikzlibrary{calc}
\usepackage{pgfplots}
\usepackage[normalem]{ulem}

\newcommand{\bea}{\begin{eqnarray}}
\newcommand{\eea}{\end{eqnarray}}

\newcommand{\bsigma}{\boldsymbol{\sigma}}

\newcommand{\br}{\mathbf{r}}

\newcommand{\bfM}{\mathbf{M}}

\newcommand{\td}{\text{d}}

\newcommand{\be}{\begin{equation}}
\newcommand{\ee}{\end{equation}}
\newcommand{\bk}{\mathbf{k}}

\newcommand{\bq}{{\bf{q}}}

\newcommand{\beal}{\begin{align}}
\newcommand{\eeal}{\end{align}}

\newcommand{\I}{\text{i}}

\begin{document}


\title{Domain wall skew scattering in ferromagnetic Weyl metals}
\author{Sopheak Sorn}
\author{Arun Paramekanti}
\affiliation{%
Department of Physics, University of Toronto, Toronto, Ontario M5S 1A7, Canada
}%

\date{\today}

\begin{abstract}
We study transport in the presence of magnetic domain walls (DWs) in a lattice model of ferromagnetic type-I Weyl metals. We compute the 
diagonal and Hall conductivities in the presence of a DW, using both Kubo and Landauer formalisms, and uncover the effect of DW
scattering. When the Fermi level lies near Weyl points, we find a strong skew scattering at the DW which leads to a significant additional
Hall effect. We estimate the average Hall resistivity for multi-domain configurations and identify the limit where the DW scattering 
contribution becomes significant. We show that a continuum model obtained by linearizing the lattice
dispersion around the Weyl points does not correctly capture this DW physics. Going beyond the linearized theory, and incorporating leading
curvature terms, leads to a semi-quantitative agreement with our lattice model results. Our results are potentially relevant for the Hall 
resistivity of spin-orbit coupled ferromagnetic metals, such as Co$_3$Sn$_2$S$_2$, Co$_2$MnGa, and SrRuO$_3$, which can have Weyl points near the Fermi energy.
\end{abstract}


\maketitle

\section{Introduction}
The anomalous Hall effect (AHE), a spontaneous deflection of electronic currents in magnetic solids, is now 
well-understood to result from two mechanisms: an intrinsic effect due to the Berry curvature of
electronic bands, and an extrinsic effect arising from impurity scattering of electrons near the Fermi level \cite{Nagaosa2010}. 
The intrinsic Berry curvature is also intimately tied to band topology and topological invariants \cite{TKNN}, as known from the two-dimensional (2D)
quantum Hall effect, where the Hall conductivity $\sigma_{xy}$ takes on a quantized value determined by the Chern 
number \cite{Zhang1989, Zhang1992, Tong2016}.  In 3D, a layered quantum Hall state with a full bulk gap can undergo a transition
into a topological Weyl semimetal as we increase the interlayer hopping \cite{Burkov2011}. The 
simplest inversion-symmetric and time-reversal broken Weyl semimetal features electronic bands which touch at two Weyl 
points \cite{Armitage2018}, around which the dispersion is approximately linear. Such a pair of Weyl points cannot be removed by any small 
perturbations, and they act as a source and a sink of the Berry curvature. When the Fermi level coincides with the energy of the Weyl points,
it leads to an intrinsic Hall conductivity $\sigma_{xy} \!=\! e^2 Q/2 \pi h$ 
where $Q$ is the momentum-space separation between the Weyl points \cite{Burkov2011}.
In fact, as a result of the linear dispersion around the Weyl points, $\sigma_{xy}$ is pinned to this value for a finite 
range of the Fermi energy around the Weyl point energy
\cite{Burkov2014}. In this regime, the system is a Weyl metal with Fermi surfaces enclosing the individual Weyl points \cite{Burkov2014}.

It is worth emphasizing that breaking time reversal  symmetry alone does not guarantee a non-zero AHE, 
even if it does lead to Weyl nodes in the band dispersion. Indeed, the
antiferromagnetic all-in-all-out ordered Weyl semimetal proposed in the pyrochlore iridates \cite{pyrochlore} is an illustrative example where a non-symmorphic glide symmetry, a mirror $M_x$ followed by a non-Bravais translation, results in a vanishing AHE. Application of uniaxial pressure on the pyrochlore iridates which
breaks this glide symmetry can then induce a non-zero AHE \cite{uniaxial_pressure_theory}.

The large AHE in several magnetic metals, including ferromagnetic
Co$_3$Sn$_2$S$_2$ \cite{ahe_c3s2s2, ahe_c3s2s2_2} and Co$_2$MnGa \cite{ahe_c2mg}, and antiferromagnetic
Mn$_3$X (X = Sn, Ge)\cite{Mn3Sn1, Mn3Sn2}, has been attributed to Weyl points in their band dispersions.
Among oxide ferromagnets, previous work \cite{Fang2003, Burkov2013, Itoh2016, Takiguchi2020} have suggested that SrRuO$_3$ \cite{rmpsro} hosts Weyl points near the Fermi level, which could account for the unusual nonmonotonic dependence of its AHE on the magnetization, including a sign-change at a certain 
temperature below $T_c$. This 
non-monotonic AHE may be understood from the magnetization dependence of the band structure, with the Weyl points and Berry curvature being 
tuned by the temperature-dependent magnetization \cite{Fang2003, Nagaosa2010, Burkov2013}. 

Remarkably, recent Hall resistivity measurements of SrRuO$_3$  thin films have discovered highly unusual hysteresis loops, with bump-like anomalies in $\rho_{xy}$ 
near the coercive field where the
magnetization begins to reverse direction as we go through the hysteresis loop \cite{Matsuno2016}. 
The origin of these anomalies is still actively debated. 
Early proposals 
regarded these bumps as an extra Hall effect induced by chiral magnetic skyrmions \cite{Matsuno2016, Pang2017, Ohuchi2018, Noh2018, Qin2019} which 
can nucleate during the magnetization reversal and can be stabilized by the interfacial Dzyaloshinkii-Moriya (DM) interactions stemming from the strong spin-orbit coupling and the inversion-breaking substrate-film interfaces \cite{Matsuno2016}. An alternative proposal argued that these anomalies emerged from imperfections in the thin films due to thickness inhomogeneities or site vacancies \cite{Kan2018, Gerber2018, Noh2020, Malsch2020, Kim2020, multichannel_new}, leading to multiple regions in space with distinct electronic and magnetic properties. Simply adding up contributions to $\rho_{xy}$ from distinct regions was argued to qualitatively reproduce the Hall anomalies
 \cite{Kan2018, Gerber2018, Noh2020, Malsch2020, Kim2020}.
  
Strikingly, measurements of the magneto-optical Kerr effect in SrRuO$_3$ films \cite{Bartram2020} discovered similar bump-like anomalies, but in films 
which were hundreds of unit cells thick, so that interfacial DM interactions and skyrmions 
play no role.
In previous theoretical work, we have shown that such anomalies in the Kerr 
effect could be captured by locally averaging the Kerr effect over 
magnetic domains \cite{Bartram2020}, an approach  justified by the locality of the high frequency response.

In contrast to our theory for the Kerr anomalies,
it is far from clear that previous theories for the Hall anomalies, which simply add up $\rho_{xy}$ from spatially distinct regions, provide a 
meaningful way to account for d.c. transport. In particular, such approaches do not explicitly account for bulk states scattering off DWs.
Given the large number of magnetic solids with Weyl points, and the ubiquity of magnetic domains in such 
systems, it is  clearly important to understand how magnetic DWs impact the Hall response of Weyl semimetals and metals. This is the key goal of our paper.

\begin{figure}[tbh]
\centering
\includegraphics[width = 0.4\textwidth]{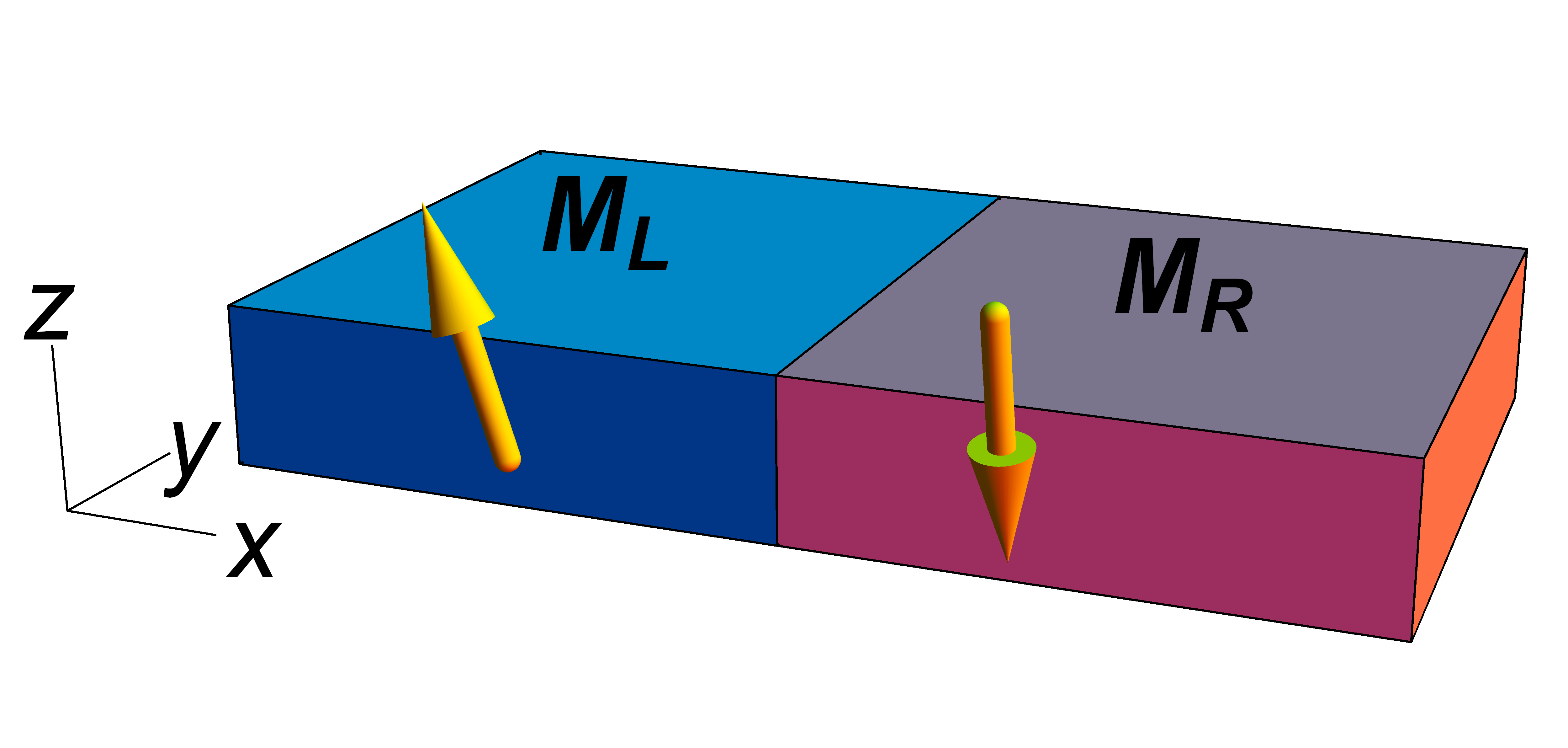}
\caption{Thin film with two magnetic domains, having uniform magnetizations $({\mathbf M}_L, {\mathbf M}_R)$, separated by a $yz$ DW.}
\label{fig:domain}
\end{figure}

In order to examine the impact of magnetic DWs on transport in a Weyl metal, we study a minimal cubic-lattice model of a ferromagnet which supports two
Weyl points in the bulk band structure. In our paper, we use the terminology `Weyl metal' as defined as in Ref. \onlinecite{Burkov2014, Burkov2018};
we use this term to refer to a system with Fermi surfaces surrounding isolated Weyl points and thus carrying nontrivial Chern number. 
The present model does not accommodate cases with additional Fermi surfaces dissociated with Weyl nodes. However, our computation of AHE from DW can be straightforwardly generalized to those cases.
Fig.~\ref{fig:domain} shows a configuration with two magnetic domains having uniform vector magnetizations
${\bf M}_L$ and ${\bf M}_R$. We assume the magnetization in each domain is uniform and choose the DW
to be in the $yz$-plane. For large domains with linear dimension much larger than the electron mean free path, we may also view such an idealized flat DW
as a section of a realistic meandering DW. In this paper, we compute the diagonal and Hall conductivities in the presence of such a DW
using a full real-space Kubo formula and compare this with a Landauer theory framework which focuses on the states near the Fermi
level scattering off the DW.  This comparison allows us to discover a strong skew-scattering contribution to the Hall transport 
arising at the DW, which is significant when the Fermi energy is not too far from the Weyl points. 

Previous theoretical work on the AHE in antiferromagnetic Weyl metal Mn$_3$Sn/Ge \cite{Liu2017} has studied Hall transport in the plane of a
magnetic DW
and shown that chiral Fermi arc modes localized on the DW can dominate this Hall effect. By contrast, our work here examines
transport in the plane perpendicular to the DW and the DW scattering of bulk states at the Fermi level. We compare our lattice model result with a continuum
theory where we linearize around the Weyl points and discover that such a linearized description completely fails to account for the lattice model calculations. 
We show that going beyond the linearized theory and incorporating leading curvature terms lead to semi-quantitative agreement with our lattice model results.
In addition to ferromagnets such as SrRuO$_3$, our results may also be broadly applicable to the AHE anomaly in
antiferromagnetic Weyl metals such as CeAlGe observed during a domain proliferation process \cite{Suzuki2019}.

This paper is organized as follows. In Section \ref{sec:model}, we introduce the lattice model for ferromagnetic Weyl metals 
and study its Hall conductivity for a uniform magnetization. In Section \ref{sec:kubo}, we consider the DW as shown in Fig.~\ref{fig:domain}
and study its impact on AHE using the real-space Kubo formula. The scattering contribution is crudely extracted and is found to be of the order of 
the Berry curvature contribution. In Section \ref{sec:dw_scattering}, 
we confirm this by extracting the DW scattering contribution using Landauer formula. Reflection coefficients (RCs) and transmission coefficients (TCs) for the Bloch states at the Fermi level scattering at the DW are found to be highly skew. We estimate the DW scattering contribution for a multi-domain configuration with 
parallel DWs and compare it with the bulk Hall contribution. In Section \ref{sec:continuum}, we linearize the lattice model around the Weyl points and 
show that RCs and TCs obtained from the linearized model lead to an incorrect result for the scattering contribution. We show that curvature 
terms are needed to reproduce the qualitative features of RCs and TCs of the lattice model. Section \ref{sec:conclusion} presents a summary and discussion.

\section{Model for Weyl metal}
\label{sec:model}
\begin{figure}[t]
\centering
\includegraphics[width = 0.5\textwidth]{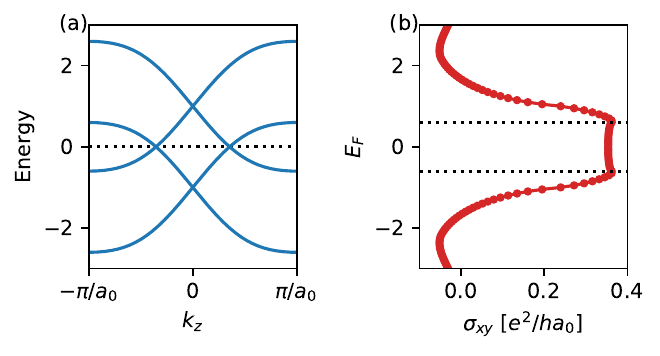}
\caption{(a) Band structure for $M = 1$ where $k_x$ and $k_y$ have been set to zero. There are two Weyl points at zero energy and at $k_z \approx \pm 1.1 a_0^{-1}$. (b) Fermi energy $E_F$ dependence of $\sigma_{xy}$ exhibiting a plateau-like behaviour, enclosed by the dashed lines, in the Weyl metal regime where the Fermi surface consists of two disjointed closed surfaces surrounding the Weyl points, and the dispersions are approximately linear.}
\label{fig:band}
\end{figure}

We consider a four-band ferromagnetic model on a cubic lattice with a uniform magnetization $\mathbf{M}$ \cite{Araki2018}:
\bea
\mathcal{H}(\bk, \mathbf{M}) &=& t(\sin k_x \sigma_x + \sin k_y \sigma_y + \sin k_z \sigma_z) \tau_z \nonumber\\ && + m(\bk) \tau
_x - J \mathbf{M}\cdot \bsigma,
\label{eq:lattice_model}
\eea
where the Hamiltonian $H = \sum_{\bk} C^{\dagger}_{\bk} \mathcal{H}(\bk, \mathbf{M}) C_{\bk}$ is defined in the basis of $C_{\bk a}^{\dagger} = \left(c^{\dagger}_{\bk A\uparrow}, c^{\dagger}_{\bk A\downarrow}, c^{\dagger}_{\bk B\uparrow}, c^{\dagger}_{\bk B\downarrow}\right)$. The Pauli matrices $\tau$ 
act on the orbital index $A \text{ and }B$, while the Pauli matrices $\sigma$ act on spin $(\uparrow,\downarrow)$. The mass term is given by
$m(\bk) = r (3 - \cos k_x - \cos k_y - \cos k_z)$.  
Time reversal symmetry is broken by the magnetization $\bfM$. For $\bfM = M \hat{z}$, the model has a four-fold rotation symmetry around the z-axis and the inversion symmetry $\tau_x \mathcal{H}(- \bk) \tau_x = \mathcal{H}(\bk)$. The dispersion is then given by 
\bea
E(\bk) &=& \pm \sqrt{t^2 (\sin^2\! k_x +\sin^2 \! k_y) + (JM \pm D(\bk))^2}, \\
D(\bk)&\equiv&\sqrt{m^2(\bk) + t^2 \sin^2\! k_z}.
\eea
For $M\!=\!0$, the band structure has a four-fold degenerate Dirac node at the $\Gamma$ point of the Brillouin zone (BZ).
With a nonzero $M$, this Dirac point splits into two Weyl points, which are located at zero energy and momenta
$\bk_{wp} \!=\! (0, 0, \pm k_z^*)$, where 
\be
\cos k_z^* \!=\! \frac{r^2 - \sqrt{t^4 + (r^2-t^2) J^2M^2}}{r^2 - t^2}.
\label{eq:kz}
\ee
The Weyl point separation $2 k^*_z$ depends on the magnetization $M$.
Figure \ref{fig:band}(a) shows the band structure for $M \!=\! 1$. In this plot, and the rest of the paper,
we fix $r \!=\! 0.8t, \text { and }J \!=\! t$. As we increase $M$, the two Weyl points move away from each other and mutually annihilate at the BZ boundary. This
results in a fully gapped quantum Hall insulator with a quantized $\sigma_{xy} = e^2G/2\pi h$ at half filling, where $G = 2\pi/a_0$ is the reciprocal lattice constant, and $a_0$ is the lattice constant of the cubic crystal. In the rest of this work, we study this model in the Weyl metal regime.

For a spatially uniform magnetization, $\sigma_{xy}$ is obtained from the momentum-space integration of the Berry curvatures
\be
\sigma_{xy} = \frac{e^2}{\hbar}\int \frac{\td^3k}{(2\pi)^3}\sum_n f(E_{\bk n})\Omega^z(\bk n),
\ee
where $f$ is the Fermi-Dirac distribution at temperature $T$, 
and $\Omega^z(\bk n)$ is the z-component of the Berry curvature vector for a state with momentum $\bk$ and band index $n$.
Figure \ref{fig:band}(b) shows $\sigma_{xy}$ at $T\!=\! 0$ as a function of Fermi energy $E_F$ for a uniform $z$-magnetization
$M \!=\! 1$. It exhibits a plateau-like behaviour in the window sandwiched between the two dashed lines, which has been studied in 
Ref.~\onlinecite{Burkov2014}. This is referred to as the Weyl metal regime where the Fermi surface consists of two disjoint closed surfaces 
surrounding the individual Weyl points, and the dispersions are approximately linear near the Weyl points. The magnitude of $\sigma_{xy}$ 
for the plateau is determined by its value at $E_F\!=\!0$, which is proportional to the momentum-space separation between the two 
Weyl points, $Q \! \equiv \! 2 k^*_z$.
For $M \!=\! 1$, the separation $Q \!=\! 2.2a_0^{-1}$, and the plateau value is given by $\sigma_{xy} \!=\! e^2 Q/2 \pi h \! \approx \! 0.35 \ e^2/ha_0 $, 
which can also be seen from Fig.\ref{fig:band}(b).

\section{Domain Wall and Hall Conductivity: Kubo formula result}
\label{sec:kubo}
We introduce a flat DW parallel to the $yz$-plane as shown in Fig.~\ref{fig:domain}, which
partitions the system into left  and right domains whose magnetizations are respectively 
denoted by $\bfM_L$ and $\bfM_R$.
Such a DW can be viewed as a locally flat region of a realistic meandering DW generated by domain proliferation 
during a magnetization reversal process in a field-sweep experiment. 
This physical picture may be a valid in the limit where the electron mean free path is much smaller than the linear 
dimensions of the domains, so that we can zoom in on electrons scattering off a small section of the domain wall.
Since the Weyl points in the minimal model Eq.~\ref{eq:lattice_model} are always 
pinned to zero energy, completely independent of the magnetization, 
we supplement this model with a term $H_{\Delta}$ that also tunes the energy of the Weyl points in 
the right domain relative to those in the left domain.
\bea
H_{\Delta} &=& \Delta \sum_{i} \Theta(i_x) ~C^{\dagger}_iC_i,
\eea
where $\Theta(i_x)$ is the lattice Heavyside step function, namely $\Theta(i_x) \!=\! 0$ for $i_x \!<\! 0$ and $1$ otherwise, and $i_x$ is the 
$x$-coordinate of the site $i$. The reason for including this term is that we envision that in a realistic setting and in material-specific models, there will be 
a relative energy shift of the Weyl points between the two domains. For instance, when domains are nucleated as we traverse the hysteresis loop in a field-sweep experiment,
this energy shift $\Delta$ could reflect a difference in the magnitude of the magnetization between majority and minority domains in the
presence of the external field, or it could reflect a local difference in the environment as
minority magnetic domains are nucleated in regions with distinct strain fields or doping or site vacancies \cite{Kim2020}. We note that such 
disorder effects by themselves, even in the absence of DWs, have been shown to have a dramatic impact for energies very close to the Weyl nodes \cite{disorder1, disorder2, disorder3}. Since our results below focuses on Weyl metals where the Fermi energy is not extremely close to the Weyl nodes, 
we expect our results on DW scattering contribution to be robust.

\begin{figure}[t]
\centering
\includegraphics[width=0.4\textwidth]{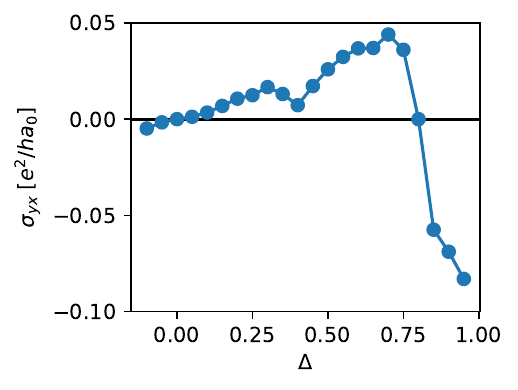}
\caption{Energy shift $\Delta$ dependence of the Hall conductivity obtained from Kubo formula in the presence of a DW. As discussed in the text, $\sigma_{yx}$ here can be regarded as purely the DW scattering contribution, which are significant and can be of the same order of magnitude as the uniform Hall conductivity $0.35e^2/ha_0$ in Fig.\ref{fig:band}(b).}
\label{fig:kubo}
\end{figure}

We compute the AHE of the above domain configuration using the Kubo formula \cite{Mahan2000, Coleman2015} (see also Appendix \ref{app:kubo}.) 
This full result contains contributions from bulk intrinsic Berry curvature as well as DW scattering effects. 
To study this, we consider a system with open boundary conditions in the $x$-direction and periodic boundary conditions 
along $y$- and $z$-directions. We choose the magnetizations to be 
$\bfM_L = M \hat{z}$ and $\bfM_R = - M \hat{z}$. Later, in Section \ref{sec:tilting}, we will discuss the effect of tilting the magnetization vector.
To obtain the AHE, which is time-reversal odd, we compute the transverse response
for a magnetic configuration and its time-reversed counterpart and subtract one from the other in order to antisymmetrize.

Fig.~\ref{fig:kubo} shows the anomalous Hall conductivity $\sigma_{yx}$ as a function of $\Delta$. Here, we have fixed the Hamiltonian parameters
$M \!=\! 1$, $J\!=\!1$ and $r\!=\!0.8 t$. We chose $E_F \!=\! 0.4t$, 
and used a system size $(L_x,L_y,L_z) \!=\! (150,300,300) a_0$, with the DW in the center at $x\!=\!L_x/2$.
As we vary $\Delta$ within the window shown in Fig.~\ref{fig:kubo}, the bulk contribution $\sigma_{yx}^{L, R}$ from deep within the interior of each domain stay roughly constant due to the plateau feature discussed in section \ref{sec:model}, and they are opposite to each other $\sigma_{yx}^L \approx - \sigma_{yx}^R$. We
thus expect the bulk contributions to nearly cancel, leaving a DW contribution $\sigma_{yx}^{DW}$ to dominate the Hall response.
Interestingly, we observe a significant contribution from the DW scattering in the limit
when the left domain is electron-like and the right domain is hole-like, i.e. $\Delta > E_F = 0.4t$.  
It can even have a similar order of magnitude as the bulk value $0.35e^2/ha_0$, e.g. at $\Delta = 0.9t$. This implies a non-negligible 
DW scattering contribution to the AHE in the Weyl metal. We now turn to study the impact of DW scattering using Landauer
theory and show that it indeed accounts for the $\Delta$-dependence of the Hall response.

\section{Domain Wall Scattering}
\label{sec:dw_scattering}

In this section, we focus on the DW scattering of bulk Bloch eigenstates, which will be used to later extract the Hall response 
using the Landauer formula \cite{Datta1995, Nazarov2009}.
We show that the transmission and the reflection at the DW exhibit a skewness, similar to the impurity-induced skew scattering in
a spin-orbit coupled ferromagnet. A notable feature is that the skewness is very pronounced when there are Weyl points near the Fermi level, which results in a significant Hall effect contribution. 
We later compare the DW scattering contribution to the bulk contribution. 
Finally, we will discuss the impact of tilting $\bfM_R$ relative to $\bfM_L$ on the Hall effect. 

\subsection{Scattering states}
In the presence of a DW in the $yz$-plane,
the eigenstates of the inhomogeneous problem $H \!+\! H_{\Delta}$ consist of bound states and scattering states. Boundary states such as Fermi arc modes, originating from a change in topology across the DW when the magnetizations are opposite, exist as bound states at the DW \cite{Ashvin2016, Araki2016, Araki2018}. Scattering states, on the other hand, are propagating waves and extend over the system.
We will focus on the scattering states which are important for studying transport across the DW. The scattering states are divided into two groups: left-incident and right-incident, denoted by $\ket{\Psi_{D; E\bk_{\parallel} \alpha}} \!=\! \Psi_{D; E\bk_{\parallel} \alpha}^{\dagger}\ket{0}$, where $D \!=\! L, R$ are the label of left- or right-incident respectively. The energy $E$ and the parallel momentum $\bk_{\parallel} \!=\! (k_y, k_z)$ are conserved quantities for elastic scattering.
 $\alpha$ is an additional label for multiple left(right)-incident channels. In the case that we will consider, there is a single incident channel once $D, E,\bk_\parallel$ are fixed
but we retain this label $\alpha$ for generality.

\begin{figure}[t]
\centering
\includegraphics[width=0.48\textwidth]{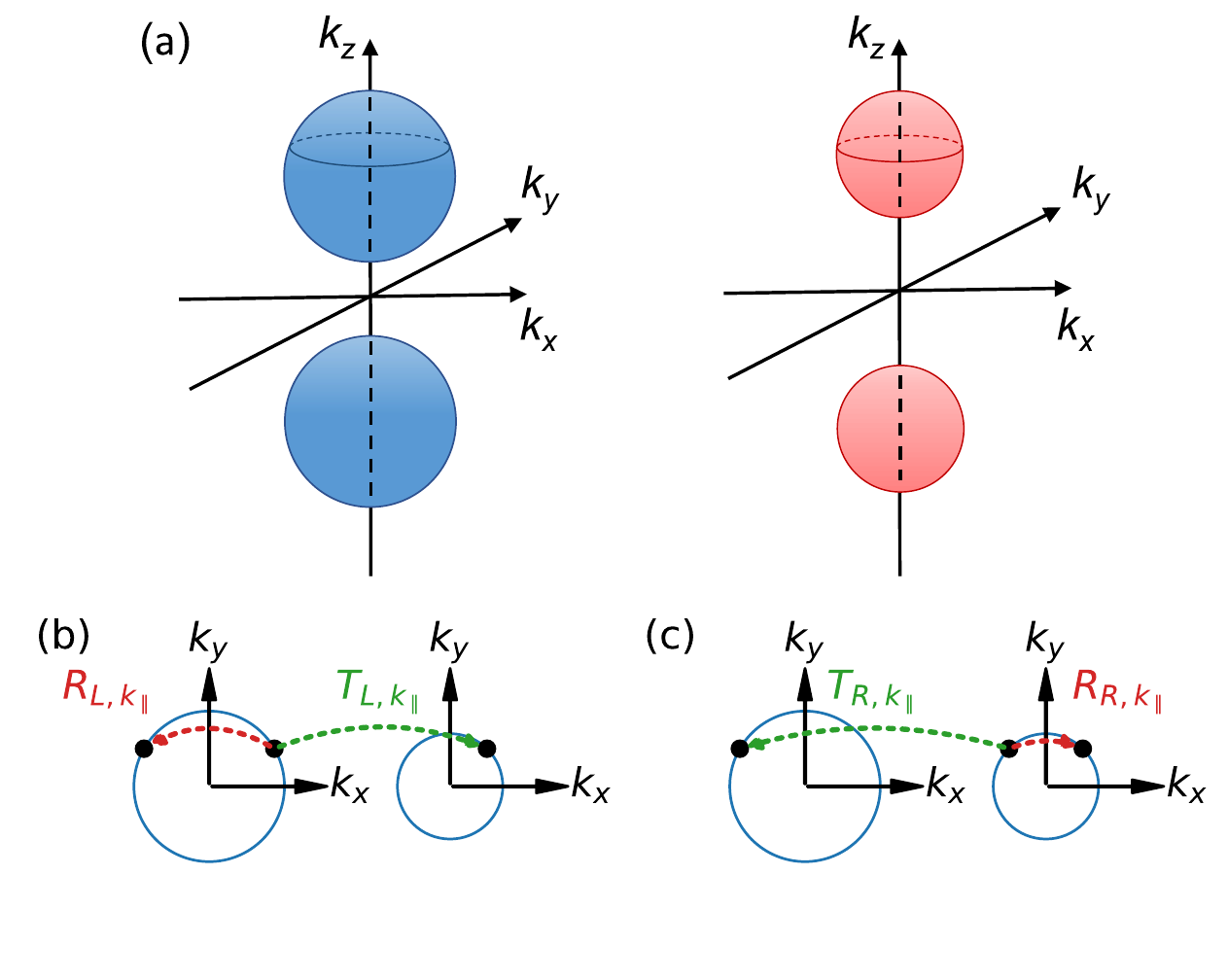}
\caption{Scattering between domains separated by a $yz$ DW as shown in Fig.~\ref{fig:domain}. The magnetizations in the two domains are chosen to be
$\pm M \hat{z}$, for which we schematically illustrate the momentum space picture of Fermi surfaces in the left and right domains in (a).
These Fermi surfaces surround the Weyl points which are separated along the $k_z$-direction due to having the magnetizations in the z-direction. 
Contours depict Fermi surface slices at a fixed $k_z$ which are used in panels (b) and (c).
(b)-(c) Schematic illustrations of the transmission coefficients $T$'s and reflection coefficients $R$'s, which connect eigenstates of 
$\mathcal{H}(\bk,\bfM_{L, R})$, for left-incident scattering states (panel (b)) and right-incident scattering states (panel (c)) at the Fermi
energy and for a fixed $k_z$.}
\label{fig:fig1}
\end{figure}

We focus on states at the Fermi level $E \!=\! E_F$.
The creation operators $\Psi_{D; E_F\bk_{\parallel} \alpha}^{\dagger}$ can be expressed in terms of the basis $C_{ia}^{\dagger}$ as the following.
\bea
\Psi^{\dagger}_{D; E_F \bk_{\parallel} \alpha} &=& \sum_{ia} \psi_{D; E_F \bk_{\parallel} \alpha} (ia) C^{\dagger}_{ia},
\eea
where $\psi_{D; E_F\bk_{\parallel} \alpha} (ia)$ is the amplitude of the scattering state at site $i$ and combined
orbital-spin label $a$. Similar to a continuum inhomogeneous problem, e.g. a potential-step problem, the amplitude can be expressed as a linear combination of the amplitudes of Bloch states of the homogeneous systems $\mathcal{H}(\bk, \bfM_{L,R})$. The coefficients of the linear combination relation can be identified with the RC and TC of the incident mode upon scattering at the DW. For instance, the amplitude of the left-incident scattering state can be written as
\bea
\!\!\!\! \psi_{L; E_F\bk_{\parallel} \alpha}\!(ia) \!\! &= \!\!
\begin{cases}
\varphi_{\alpha\bk_{\parallel}} \! (ia) \! + \! \sum\limits_{\beta} r^{\alpha \rightarrow \beta}_{L; \bk_{\parallel}} \varphi_{\beta\bk_{\parallel}}\! (ia) & \!\!\!\! (i_x \! < \! 0),\\[12pt]
\sum\limits_{\tilde{\beta}} t^{\alpha \rightarrow \tilde{\beta}}_{L;\bk_{\parallel}}\tilde{\varphi}_{\tilde{\beta}\bk_{\parallel}}\! (ia) & \!\!\!\! (i_x \! \ge \! 0),
\end{cases}
\label{eq:left_incident}
\eea
where $\varphi_{\alpha\bk_{\parallel}}(ia)$ and $\varphi_{\beta \bk_{\parallel}}(ia)$ 
are the amplitudes of the incident and reflected Bloch waves respectively, which are eigenstates of 
$\mathcal{H}(\bk, \bfM_L)$. $\tilde{\varphi}_{\tilde{\beta} \bk_{\parallel}} (ia)$ 
is the amplitude of a transmitted Bloch wave, which is an eigenstate of $\mathcal{H}(\bk, \bfM_R)$.
These are the states at the Fermi surfaces surrounding Weyl points as shown schematically in Fig. \ref{fig:fig1}(a).
The $\beta$ summation is performed over all the reflected channels, whereas the $\tilde{\beta}$ summation is carried over all the transmitted channels. These also include evanescent waves which are eigenstates of $\mathcal{H}(\bk, \bfM_{L, R})$ corresponding to complex-valued $k_x$ and decay exponential away from the DW.
RC for the incident mode $\alpha$ going into a reflected mode $\beta$ is given by $R_{L, \bk_{\parallel}}^{\alpha \rightarrow \beta} = |r_{L, \bk_{\parallel}}^{\alpha \rightarrow \beta}|^2 |v_{x, \beta}/v_{x, \alpha}|$ where $v_{x, \alpha}$ and $v_{x, \beta}$ are their group velocities in the x-direction. Similarly, TC is defined by $T^{\alpha \rightarrow \tilde{\beta}}_{L; \bk_{\parallel}} = |t^{\alpha \rightarrow \tilde{\beta}}_{L; \bk_{\parallel}}|^2 |v_{x, \tilde{\beta}}/v_{x, \alpha}|$. Figure \ref{fig:fig1}(b) illustrates how the eigenstates of the homogeneous problems $\mathcal{H}(\bk, \bfM_{L, R})$ on their Fermi surfaces 
are connected by RCs and TCs in the left-incident scattering states.

Similarly, the right-incident scattering states can be constructed from the eigenstates of the homogeneous problems as the following.
\bea
\!\!\!\! \psi_{R; E_F \bk_{\parallel}, \tilde{\alpha} } \! (ia) \!\! &= \!\! \begin{cases}
\sum\limits_{\beta} t^{\tilde{\alpha} \rightarrow \beta}_{R; \bk_{\parallel}} \varphi_{\beta, \bk_{\parallel}}\!(ia) & \!\!\!\!\!\! (i_x \!\! < \! 0),\\[12pt]
\tilde{\varphi}_{R;\tilde{\alpha}, \bk_{\parallel}} \! (ia) \! + \!\! \sum\limits_{\tilde{\beta}} r^{\tilde{\alpha} \rightarrow \tilde{\beta}}_{R;\bk_{\parallel}}\tilde{\varphi}_{\tilde{\beta},\bk_{\parallel}} \!\! (ia) & \!\!\!\!\!\! (i_x \!\! \ge \! 0).
\end{cases}
\label{eq:right_incident}
\eea
TCs and RCs for the right-incident scattering states are illustrated schematically in Fig.\ref{fig:fig1}(c).
TCs and RCs for the inhomogeneous lattice model are computed using a method described in great detail in 
Ref. \onlinecite{Ando1991}.

\subsection{Skew reflection and skew transmission}
\label{sec:skew}
\begin{figure}
\centering
\includegraphics[width=0.4\textwidth]{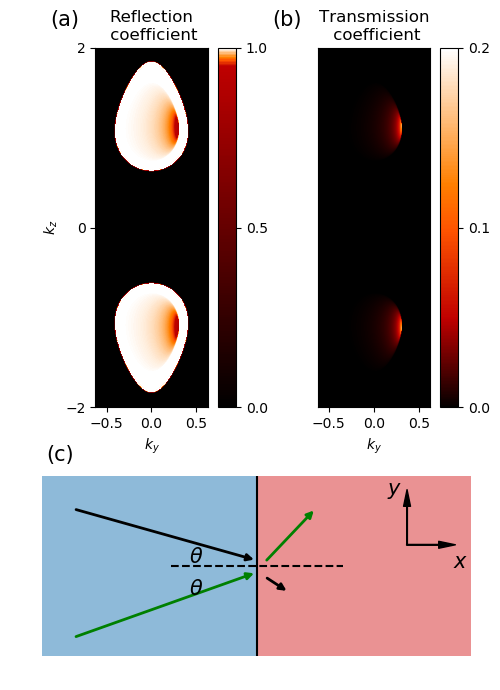}
\caption{(a) Reflection coefficient and (b) transmission coefficient for left-incident scattering states as a function of their labels $\bk_{\parallel}$, featuring a pronounced skewness between any pair of state with the same $k_z$ and opposite $k_y$. This implies that a pair of incident Bloch waves moving opposite to each other in the y-direction get transmitted asymmetrically at the DW, as illustrated in (c). This results in a transverse Hall current in the presence of a bias voltage between the two domains.}
\label{fig:fig2}
\end{figure}

Figure \ref{fig:fig2} shows RCs and TCs for the left-incident scattering states as a function of $\bk_{\parallel} = (k_y, k_z)$ for $\Delta = 0.1t$, which is one of the cases studied in Section \ref{sec:kubo}. At this Fermi energy, there are at most one incident, one reflected and one transmitted channels (excluding evanescent channels which do not participate in transport.) Therefore, there are only one TC and one RC for each scattering state.
We have checked that the RC and TC  sum to unity for every $\bk_{\parallel}$. The dark regions in (a), where both RC and TC are zero, correspond to the regions where there are no incident modes. We observe that TCs are highly skew between every pair of scattering states whose $k_y$ momenta are opposite. Namely, at a fixed $k_z$, TC for a positive $k_y$ is large, while that for $-k_y$ is extremely small. 
This feature is observed for the entire range of $\Delta$ studied in Section \ref{sec:kubo}.

Such highly skew features result in a large Hall effect, which can be seen by considering any pair of left-incident scattering states whose $k_y$ momenta are opposite, i.e. $(k_y, k_z)$ and $(-k_y, k_z)$. The two incident Bloch states move with the opposite group velocities in the $y$-direction and get transmitted asymmetrically at the DW, as illustrated in Fig.\ref{fig:fig2}(c). This produces a transverse Hall current when a bias voltage between the two domains is applied. This will be studied quantitatively in the next subsection by computing Hall conductance using Landauer formula. Such skewness is, in fact, expected in systems with strong spin-orbit couplings \cite{Fabian2015, Zhuravlev2018}. However, our new result here is that the skewness is very pronounced when the Fermi level resides near the Weyl points. We have checked that when $E_F$ is far away from the Weyl points, such skewness is weaker, and TCs are many order of magnitude smaller (see Appendix \ref{app:A}), which leads to a very small Hall contribution. Thus, DW scattering is significant when there are Weyl points near $E_F$.

\subsection{Landauer theory of domain wall Hall conductance}
\label{sec:dw_conductance}
Hall conductance arising from DW scattering and longitudinal conductance across the DW can be computed within the Landauer formalism \cite{Datta1995, Nazarov2009} using TCs and RCs obtained above. In the presence of an applied bias voltage $\Delta V_x$, a current density along x-direction $j_x$ and y-direction $j_y$ are produced. These can be computed using the scattering states as derived in Appendix \ref{app:B} (see also Ref. \onlinecite{Fabian2015} and Ref. \onlinecite{Zhuravlev2018}.) From these, we obtain the expressions for the conductance per unit cross section area $g_{xx} = j_x / \Delta V_x$ and $g_{yx} = j_y/\Delta V_x$ as shown below
\bea
g_{xx} \!\! &=& \!\! \frac{e^2}{h} \!\! \int \!\!
\frac{\td \bk_{\parallel}}{(2\pi)^2} \! \left[T^{\alpha \rightarrow \tilde{\beta}}_{L;\bk_{\parallel}}\right]_{E=E_F},\\
g_{yx} \!\! &=& \!\! \frac{e^2}{2h} \!\! \int \!\! \frac{\td \bk_{\parallel}}{(2\pi)^2} \!\! \left[\frac{v_{y\alpha}}{|v_{x\alpha}|} \!+\! \frac{v_{y\beta}}{|v_{x\beta}|} R^{\alpha \rightarrow \beta}_{L;\bk_{\parallel}} \! + \! \frac{v_{y\tilde{\beta}}}{|v_{x\tilde{\beta}}|} T^{\alpha \rightarrow \tilde{\beta}}_{L;\bk_{\parallel}}\right]_{E=E_F}
\eea
Summations over the incident channel $\alpha$, reflected channel $\beta, \text{ and transmitted channel } \tilde{\beta}$ are implicit. These expressions are valid at zero temperature where only states at $E_F$ are important. To obtain anomalous Hall response which is time-reversal odd, we antisymmetrize $g_{yx}$ as described in Section \ref{sec:kubo} for the Kubo Hall conductivity; we will continue to refer to the antisymmetrized version as $g_{yx}$ in the rest of the paper.

\begin{figure}
\centering
\includegraphics[width = 0.5\textwidth]{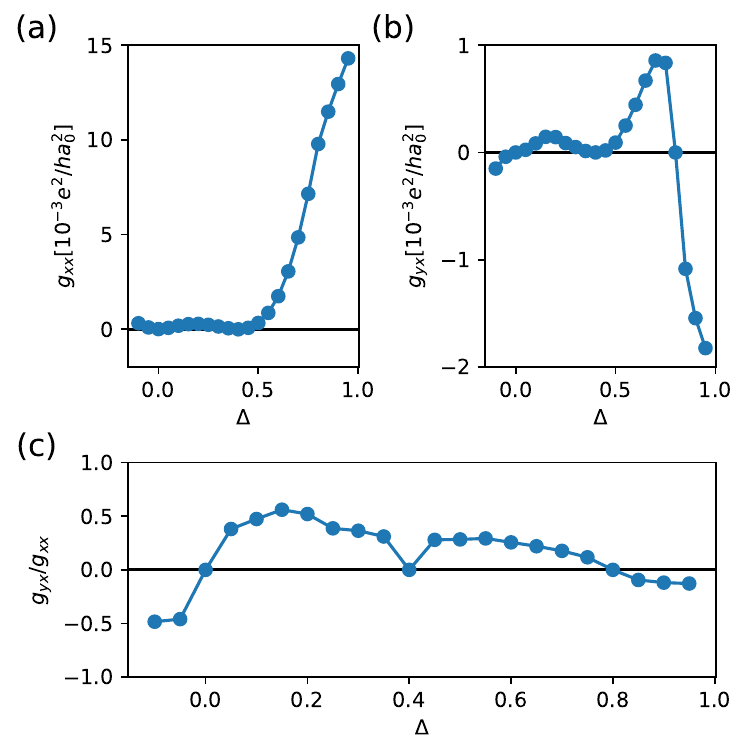}
\caption{(a) Longitudinal conductance per unit cross section area $g_{xx}$ (b) Hall conductance per unit cross section area $g_{yx}$ and (c) their ratio $g_{yx}/g_{xx}$ 
as function of the energy shift $\Delta$.}
\label{fig:fig3}
\end{figure}

Figure \ref{fig:fig3}(a) and (b) show the $\Delta$ dependence of $g_{xx}$ and $g_{yx}$ for the model parameters as in Section \ref{sec:kubo}.  
The ratio $g_{yx}/g_{xx}$ shown in Fig.\ref{fig:fig3}(c) is rather large and is of the order of 10$^{-1}$, which is a consequence of having highly skew TCs. We will show that the largeness of this ratio leads to an observable DW scattering contribution. Before that we first compare $g_{yx}$ to the Hall conductivity obtained from Kubo formula in Section \ref{sec:kubo}.

We observe that the $\Delta$ dependence of $g_{yx}$ in Fig.\ref{fig:fig3}(b) and that of $\sigma_{yx}$ in Fig.\ref{fig:kubo} bear a strong resemblance. More importantly, they have the same sign at each $\Delta$. These suggest that $\sigma_{yx}$ in Fig.\ref{fig:kubo} indeed tracks the DW scattering contribution which arises from skew scatterings at the DW. 
The connection between $g_{yx}$ and $\sigma^{DW}_{yx}$ may be established by the following argument. 

In the Kubo approach, we suppose that the Bloch electrons have a lifetime $\hbar/\gamma$, where $\gamma$ is an energy broadening used in the 
single-particle Green's function. This translates to a mean free path $\ell_0 \!=\! v_F \hbar/\gamma$, where $v_F \!=\! \partial E /\hbar \partial k$ is the Fermi velocity. 
Only electrons at a distance less than $\ell_0$ from the DW can experience DW scattering, and they see a potential drop $\Delta V_x \sim \ell_0 E_x$ across the DW, 
where $E_x$ is the electric field.  We thus infer from the Kubo calculation, a transverse current density due to DW scattering in this region, given by
$j_y = \sigma^{DW}_{yx} (\Delta V_x/\ell_0) \equiv g_{yx}^{DW}  \Delta V_x$. Thus, 
$g_{yx}^{DW} =\sigma^{DW}_{yx}/\ell_0$, which can be compared with $g_{yx}$ in the Landauer formalism. Now, we have earlier argued that the Kubo response
for our specific domain
configuration is expected to have cancelling bulk contribution, so the entire result is expected to be dominated by $\sigma_{yx}^{DW}$. We will thus use
the computed curve in Fig.~\ref{fig:kubo} as our estimate for $\sigma^{DW}_{yx}$.
Our choice of $\gamma = 0.01 t$ used in the Kubo calculation, with $v_F \!\approx\! a_0t/\hbar$ from the band structure, where $t$ is the hopping parameter and $a_0$ is the lattice constant, then leads to $\ell_0 = 100 a_0$. 
We thus  expect $g_{yx}^{DW}\! \approx\! \sigma_{yx}^{DW}/100 a_0 $. This is in reasonable agreement (within a factor of two) with the Landauer
result shown in Fig.~\ref{fig:fig3}(b).
We have also checked that increasing $\gamma$, which reduces $\ell_0$, leaves our estimated $g_{yx}^{DW}$ to be nearly unchanged, so that this
agreement between the Kubo and Landauer results is not sensitive to the choice of $\gamma$ so long as it is not too small. The finite lifetime of Bloch electron can arise from disorder in the lattice. It has been shown that 
disorder can have pronounced effects on Weyl semimetals when the Fermi level is very close to the Weyl points, e.g. driving a phase transition into a diffusive 
metallic phase \cite{disorder1, disorder2, disorder3}. However, our above results for domain wall skew scattering thus remain 
valid as long as the Fermi energy is not too close to the Weyl points.

\subsection{Multidomain configurations: comparing bulk versus DW scattering contribution}

The DW scattering contribution in a transport experiment will clearly be sensitive to the number of minority domains and their domain sizes. For few and small
minority domains, the measured Hall response will be dominated by the intrinsic bulk contribution. As we increase the number of minority domains,
the DW contribution will increase, while the net bulk contribution will decrease due to partial cancellation between majority and minority
domains.
To gain a perspective on when the DW scattering contribution becomes significant relative to the bulk intrinsic contribution, 
we consider a simple multi-domain setting with a series of parallel $yz$-DWs. 
For simplicity, let us assume two types of domains with collinear 
magnetizations, $\bfM_+ \!=\! M \hat{z}$ and $\bfM_{-} \!=\! - M \hat{z}$, pointing along the $z$-direction, and
$N_{DW}$ DWs over the sample length $L_x$, so that the average distance between two neighbouring DWs is 
$L_x/N_{DW}$.  Such a configuration has a $z$-mirror symmetry, under which
$(x,y) \!\to\! (-x,-y)$ but the  magnetizations are left invariant. This enforces the conductances $G_{yz} \!=\! G_{xz} \!=\! 0$, thus simplifying the conductance tensor.

\begin{figure}[t]
	\centering
	\includegraphics[width=0.4\textwidth]{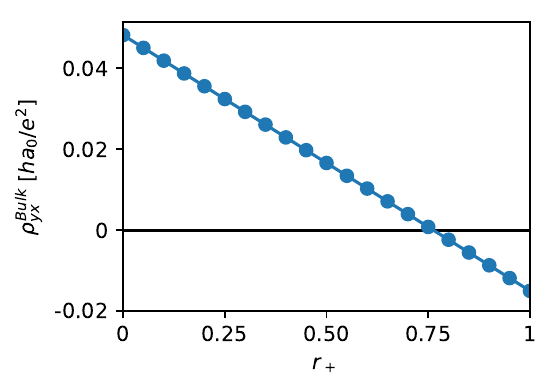}
	\caption{The dependence of the volume-averaged bulk Hall resistivity $\rho_{yx}^{Bulk}$ from Eq.\ref{eq:bulk}, 
	on the volume fraction $r_\pm$ of the magnetic domains $\mathbf{M}_\pm = \pm M \hat{z}$, 
	with $r_++r_-\!=\! 1$.
	The displayed  result corresponds to $M\!=\! 1$, $\Delta \!=\! 0.1t$ and $E_F \!=\! 0.4t.$ }
	\label{fig:estimate}
\end{figure}

In the limit where the electron mean free path $\ell_0 \ll  L_x/N_{DW}$, we 
consider an $x$-interval $(x_{DW}\!-\! \ell_0/2, x_{DW}\!+\! \ell_0/2)$, centered around a DW at $x_{DW}$, where the Hall effect may be dominated by DW scattering. 
In the presence of a current density $j_x$, the Hall voltage in this DW region is given by 
$V_{H, DW}\! =\! 
- (\sigma^{DW}_{yx}/\sigma^{DW}_{xx} \bar{\sigma}_{yy})j_x L_y$,
where the diagonal conductivity
along $y$ in the interval $(x_{DW}\!-\!\ell_0/2, x_{DW}\!+\! \ell_0/2)$
may be approximated as the average $\bar{\sigma}_{yy} \!\approx\! (\sigma_{yy}^{(+)} \!+\! \sigma_{yy}^{(-)})/2$. We thus obtain
$V_{H, DW} \!=\!  - (g_{yx}/g_{xx}\bar{\sigma}_{yy}) j_xL_y$, where $g_{yx}$ and $g_{xx}$ are the DW scattering contributions. 
Away from the DWs, the bulk Hall voltages are $V_{H, b} = \rho_{yx}^{(b)} j_x L_y$, where $b = \pm$. Using the relation
$\rho_{yx} = -\sigma_{yx} / \sigma_{xx} \sigma_{yy}$, the Hall voltage averaged over the $x$-axis is given by the expression
$V_{H, av} = (\rho_{yx}^{Bulk}+\rho_{yx}^{DW})  j_x L_y$, where
\bea
\rho_{yx}^{Bulk}&=&- \sum_{b=\pm} r_b \frac{\sigma^{(b)}_{yx}}{\sigma_{xx}^{(b)}\sigma^{(b)}_{yy}},
\label{eq:bulk}\\
\rho_{yx}^{DW}&=&-\frac{g_{yx}}{g_{xx}}\frac{1}{\bar{\sigma}_{yy}} \frac{N_{DW} \ell_0}{L_x},
\label{eq:dw}
\eea
where $r_\pm$ are the volume fraction of domain $b=\pm$. 
We have used the fact that the Hall voltage at the DW between $(\bfM_+|\bfM_-)$ is identical to that between $(\bfM_-|\bfM_+)$, which is due to the fact that 
these two DW configurations are related by a local inversion operation. This allows their contributions to add up instead of cancelling each other. 
The bulk contribution in Eq.\ref{eq:bulk} depends on $r_b$, as illustrated in Fig.\ref{fig:estimate}. Since the bulk contributions from the $\bfM_+$ and the $\bfM_-$ domains are opposite in sign, their sum becomes zero at a nonzero $r_+$, leaving the DW contribution to dominate the Hall effect. 
Near this $r_+$, we can see from Eq.\ref{eq:dw} that the DW scattering contribution to the bulk value 
DW scattering contribution continues to dominate over the bulk contribution as long as the spacing between DWs 
falls below a threshold value $\sim \ell_0 |(g_{yx}/g_{xx}\bar{\sigma}_{yy})/\rho_{yx}^{Bulk}|.$ From Fig.~\ref{fig:fig3}, $g_{yx}/g_{xx} \!\sim\! 0.5$ for $\Delta\!=\!0.1t$, and 
Fig.~\ref{fig:estimate} shows
$\rho^{Bulk}_{yx} \!\sim\! 10^{-2}$ over a range of $r_+$. Using $\bar{\sigma}_{yy}\sim 5 e^2/ha_0$ from the Kubo calculation with the corresponding $\ell_0 \approx 100a_0$, we expect the DW Hall effect, and the corresponding anomalies in the Hall transport,
to be detectable even for large DW separation  $\sim \! 10\ell_0$.
Finally, we note that the $\Delta$ dependence of TCs and the conductance densities in the case when the domain magnetizations are unequal in magnitude is qualitatively
similar to the case discussed here, as long as their magnitudes are not too different.

\subsection{Impact of tilting the magnetization vectors}
\label{sec:tilting}
In a realistic system, there can be an easy-axis anisotropy along a certain low-symmetry direction. During a magnetization reversal process where magnetic domains proliferate and under an applied magnetic field in the z-direction, the magnetizations in the majority and the minority domains can become non-collinear. We study the impact of such non-collinearity on the DW Hall effect here. Setting $\Delta$ to zero and setting the norm of magnetization to unity, we consider two tilting cases: (1) $\bfM_L = \hat{z}$ and a tilting parallel to the $yz$ DW $\bfM_R = - (0, \sin \theta, \cos\theta)$  and (2) $\bfM_L = \hat{z}$ and a tilting out of the $yz$ DW $\bfM_R = - (\sin \theta, 0, \cos\theta)$.

\begin{figure}[t]
	\centering
	\includegraphics[width=0.49\textwidth]{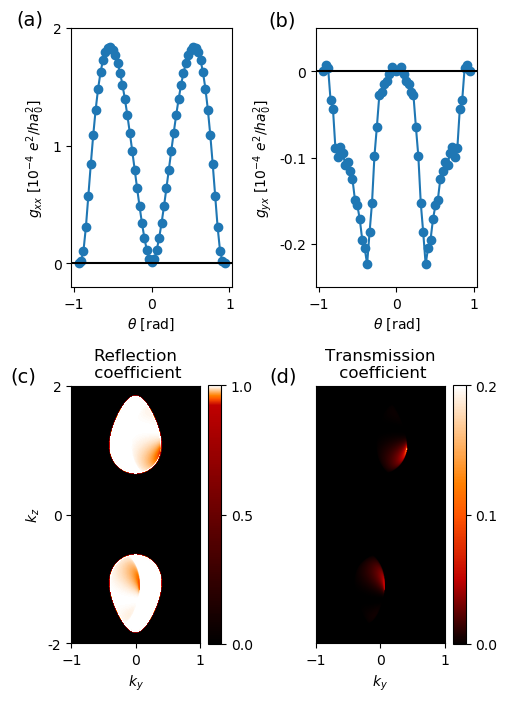}
	\caption{(a) Longitudinal conductance per unit cross section area $g_{xx}$ and (b) Hall conductance per unit cross section area $g_{yx}$ as a function of the in-plane tilting angle $\theta$. (c) Reflection coefficients and (d) transmission coefficients corresponding to the negative $\theta$ where the magnitude of $g_{yx}$ is the largest.}
	\label{fig:tilting1}
\end{figure}

Figure \ref{fig:tilting1}(a) and (b) show the impact of tilting in case (1) on the conductance per unit cross section area. $g_{yx}$ is nonzero due to the combination of (i) asymmetry of TCs as shown in Fig.\ref{fig:tilting1}(d) and (ii) the asymmetry of the magnitude of the group velocity $|v_y|$ which was absent in previous discussions and is now expected when the magnetization has a non-vanishing y-component. $g_{yx}$ is an even function of $\theta$, which can be understood by how the Hall conductivity transforms under the mirror $\mathcal{M}_y$ followed by time reversal $\mathcal{T}$. The magnitudes of both $g_{xx}$ and $g_{yx}$ are smaller than those in Fig.\ref{fig:fig3}. However, the $g_{yx}/g_{xx}$ ratio remains large and of the same order $\sim 10^{-1}$, so the DW scattering contribution to Hall effect is expected to be noticeable in experiment as inferred from Eq.\ref{eq:dw} 
\footnote{Rigorously, one needs to generalize Eq.~12 and Eq.~13, which are valid when the x-component and the y-component of the magnetization vanish, to include the nonzero $G_{yz}$ and $G_{xz}$. These have contributions from the Fermi sea (Berry curvature), the DW scattering (Fermi surface) and even Fermi arc states \cite{Liu2017}, which is beyond the scope of the Landauer formalism.}.

In case (2), $g_{yx}$ is zero for all the tilting angle $\theta$ (not shown). TCs are found to be symmetric ($|v_y|$ is also symmetric since the y-component of the magnetization vanishes.) It is unclear what protects TCs from being asymmetric. The net Hall conductivity (intrinsic + extrinsic) is non-zero in this case as allowed by symmetry, namely the broken $\mathcal{M}_y$ and the broken $\mathcal{T}$ are sufficient to allow a non-zero Hall effect in the xy-plane. However, the Hall contribution from the DW scattering vanishes. $g_{yx}$ becomes non-zero when $\Delta$ is set to non-zero. It is possible that a combination of particle-hole symmetry broken by a 
non-zero $\Delta$ and $\mathcal{M}_y \mathcal{T}$ broken by a non-zero y-component of the magnetization are responsible for the symmetric TCs. We have not
done a full symmetry analysis of the TCs; we defer this to future work.

\section{Continuum model of Weyl metal}
\label{sec:continuum}
In this section, we discuss the DW scattering within a continuum model obtained from Taylor expanding the lattice model dispersion around the Weyl points. We find that at linear order in momentum, the continuum model can lead to an incorrect result, while a qualitative agreement with the lattice model is obtained when we keep quadratic terms. This suggests that the higher order terms are important for studying DW scattering in Weyl metals.

For $M_L = M_R = 1$ corresponding to the magnetizations in the z-direction, the Weyl points reside at the same momentum positions for both domains 
$\bk_{wp} = (0, 0, \pm k_z^*)$ where $k_z^*$ is given in Eq.~\ref{eq:kz}.
Let $\bq \equiv \bk - \bk_{wp}$. The linearized continuum model is given by
\bea
\label{eq:linearized}
\mathcal{H}(\bq)_{L/R} &=& tq_x\sigma_x \!+\! tq_y \sigma_y \! \mp \! J\sigma_z \! + \! \sigma_z [t(\sin k_z^* \! + \! \cos k_z^* q_z) \tau_z  \nonumber\\
&& + r(1 \! -\! \cos k_z^* \! +\! \sin k_z^* q_z) \tau_x], 
\eea
where we have performed a unitary transformation $U = \text{diag}(1, 1, 1, -1)$ on Eq.\ref{eq:lattice_model} before the linearization. Here $q_x$ is viewed as an operator $q_x = -i\partial_x$ since the translational invariance along the x-direction is broken. $q_y$ and $q_z$ are still good quantum numbers and can be treated as numbers. We can diagonalize the term in the square bracket for a given $k_z$. As a result, the two domains can be simultaneously block diagonalized into the following form.
\bea
\mathcal{H}(\bq)_{L/R} \! &=& \! tq_x\sigma_x \! + \! tq_y \sigma_y \! + \! \sigma_z\begin{pmatrix}
 Z_+ \! \mp \! J & 0\\
0& Z_- \! \mp \! J
\end{pmatrix},
\eea
where $Z_{\pm}$ are the eigenvalues of the matrix in the square bracket in Eq.\ref{eq:linearized}. Let $Z_+ \ge 0$ and $Z_- \le 0$. In the left domain with $-J$, all propagating-wave solutions at a small, positive Fermi level reside on the bands associated with the Weyl points and thus correspond to the upper block where the mass term $Z_+-J$ can become zero. This means that the propagating-wave solutions in the left domain have \emph{zero} weight in the lower-block entry. For the right domain, the mass term $Z_- + J$ in the lower block can instead become zero. Therefore, the propagating-wave solutions in the right domain have zero weight in the \emph{upper} block. These result in a zero transmission for all $(q_y, q_z)$ since the incident modes from the left domain are orthogonal to the transmitted modes in the right domain. This result is robust against adding the energy shift term $H_{\Delta}$. Therefore, at linear order, the continuum model predicts a zero transmission and suggests an infinite DW resistance. This is obviously incorrect, for we have seen nonzero transmission and a rich $\Delta$ dependence of TC in the full lattice model. The orthogonality and the existence of a basis where $\mathcal{H}_{L/R}$ can be simultaneously block diagonalized are an artefact of the linearized model and can be removed by keeping higher order terms. 

\begin{figure}
\centering
\includegraphics[width=0.49\textwidth]{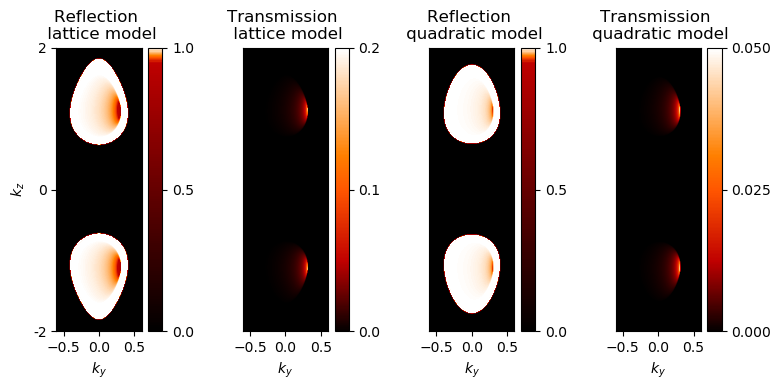}
\caption{Reflection and transmission coefficients obtained from the lattice model (left two panels) and the quadratic model (right two panels), featuring their qualitative agreement in terms of their asymmetry in $k_y$. Without the quadratic terms, the continuum model yields zero transmission (perfect reflection), as explained in the text, which would correspond to a completely dark color plot (not shown here). All these suggest that higher-order terms in momentum are responsible for the skew transmission observed in the lattice model.}
\label{fig:quadratic}
\end{figure}

Fig.~\ref{fig:quadratic} shows the comparison between the RCs and TCs obtained within the lattice model and within the 
low energy description when we include leading curvature terms by going to quadratic order in the expansion around the Weyl points.
These results are for the same model parameters as studied in Section \ref{sec:skew}.  The results for the quadratic model are obtained by 
matching both the wavefunctions and their first derivatives $\partial \psi/\partial x$ at the DW. We find reasonable semi-quantitative agreement between
the lattice and continuum descriptions at this order. Such an agreement persists for the whole range of $\Delta$. Thus, we find that going beyond
the linearized theory is important for a proper description of transport across a DW.

Finally, we note that when $M_L \neq M_R$, the linear model predicts nonzero TCs like in the lattice model. However, 
they are a few orders of magnitude smaller. All these suggest that (1) the linear model is insufficient for studying the 
DW scattering and (2) higher order terms are responsible for the results discussed in the previous sections.

\section{Conclusion}
\label{sec:conclusion}
Using a minimal model of ferromagnetic Weyl metal containing a pair of Weyl points, we have shown that DW scattering for states on the Fermi surfaces surrounding the Weyl points is highly skew. This can lead to a large, observable AHE contribution. For Fermi level away from Weyl points, the effect of DW scattering diminishes. 
Therefore, the DW scattering contribution must not be neglected when there are Weyl points near the Fermi level.
A continuum model obtained from linearizing the Weyl metal lattice model around the Weyl points fails to capture this result. We show that curvature 
terms in momentum are needed to qualitatively reproduce the results of the lattice model.

Generalization of our results to a more complicated model of Weyl semimetal or Weyl metal, e.g. with multiple pairs of Weyl points or with parasitic Fermi surfaces unrelated to any Weyl points, must proceed with care. If the DW scattering state, Eq.\ref{eq:left_incident} or Eq. \ref{eq:right_incident}, involves states from two Fermi surfaces enclosing Weyl points (one from left domain and one from right domain), our result can be directly deployed.
However, when a parasitic Fermi surface or more than two Fermi surfaces enclosing Weyl points are involved in the DW scattering states, one needs a new computation, as a straightforward generalization of the calculations presented in this paper.

Our results call for a re-examination of AHE in SrRuO$_3$ 
thin films through a realistic model in the presence of DWs in order to understand the 
peculiar bumps features in the AHE hysteresis loops. Our results may also be tested in transport experiments on ferromagnetic Weyl metals 
Co$_3$Sn$_2$S$_2$ \cite{ahe_c3s2s2, ahe_c3s2s2_2} and Co$_2$MnGa \cite{ahe_c2mg}.
Our results also suggest that the extra AHE observed in the 
antiferromagnetic Weyl metal CeAlGe during a magnetic domain proliferation process could be attributed to DW scattering \cite{Suzuki2019}.
Another important message of our work is that a careful account of such DW scattering must be taken into consideration before one can 
attribute Hall resistivity anomalies to the topological Hall effect due to skyrmion spin textures.

\begin{acknowledgments}
The authors thank Anton A. Burkov and Arijit Haldar for extremely useful discussions and suggestions. We also acknowledge an ongoing related collaboration with
Andrew Hardy. This work was funded by NSERC of Canada. This research was  enabled  in  part  by  support  provided  by  WestGrid  (www.westgrid.ca) and Compute Canada  Calcul Canada (www.computecanada.ca).
\end{acknowledgments}

\appendix

\section{Kubo formula}
\label{app:kubo}
In linear response, the d.c. conductivity is given by \cite{Mahan2000, Coleman2015}
\begin{widetext}
\bea
\label{eq:kubo_full}
\sigma_{\alpha\beta} &=& \lim_{\omega \rightarrow 0} \frac{\I 2\pi}{V}\frac{e^2}{h}\sum_{\bk_{\parallel},m,n} \frac{f(E_{\bk_{\parallel} m}) - f(E_{\bk_{\parallel} n})}{E_{\bk_{\parallel} n} - E_{\bk_{\parallel} m}} \left[\frac{(\hat{j}_{\alpha})_{mn} (\hat{j}_{\beta})_{nm}}{\hbar \omega + i \gamma + E_{\bk_{\parallel} m} - E_{\bk_{\parallel} n} }\right],
\eea
\end{widetext}
where $V$ is the volume, $f$ is the Fermi distribution function, $\bk_{\parallel} = (k_y, k_z)$, $E_{\bk_{\parallel} m}$ is an energy eigenvalue, $m, n$ are the band indices , $(\hat{j}_{\alpha})_{mn} = \bra{\bk_{\parallel} m} \hat{j}_{\alpha} \ket{\bk_{\parallel} n}$, and $\gamma$ is a small broadening. The current operator is obtained from Peierls substitution in each hopping term $t_{ia,jb}c^{\dagger}_{ia}c_{ib}$ as the following.
\bea
\!\!\!\!\! t_{ia, jb}c^{\dagger}_{ia}c_{jb} &\rightarrow & t_{ia, jb} c^{\dagger}_{ia}c_{jb} \exp \left(\I \int_i^j \td \mathbf{r} \cdot \mathbf{A}\right),\\
\!\!\!\!\! &\approx & t_{ia, jb} c^{\dagger}_{ia}c_{jb} \left(1 + \I \br_{ij}\frac{\mathbf{A}(i) + \mathbf{A}(j)}{2} \right),
\eea
where $\br_{ij} = \br_j - \br_i$, and $\mathbf{A}$ is a vector potential corresponding to an electric field $\mathbf{E} = - \partial \mathbf{A}/\partial t$. The current operator at a site $i$ is given by
\bea
\hat{\mathbf{j}}(i) &=& \frac{\delta H\left[\mathbf{A}\right]}{\delta \mathbf{A}(i)},
\eea
where $H\left[\mathbf{A}\right]$ is the Hamiltonian after the Peierls substitution. A real-space expression of the lattice model, Eq.\ref{eq:lattice_model}, can be found in Ref.\onlinecite{Araki2018}.

The current operator in Eq.\ref{eq:kubo_full} is defined by $\hat{j}_{\alpha} = \sum_i \hat{j}_{\alpha}(i)$. Since the system has translational invariance along y and z, $\hat{j}_{\alpha}$ can be Fourier transformed partially in the (y, z) space and becomes block diagonal in $\bk_{\parallel}$. Each block corresponding to a $\bk_{\parallel}$ is a $4L_x$ by $4L_x$ matrix, where $4$ is the number of band of the model in the homogeneous case. This matrix can be used to numerically evaluate $(\hat{j}_{\alpha})_{mn}$. 

\section{Transmission coefficient away from Weyl points}
\label{app:A}
\begin{figure}[h]
\centering
\includegraphics[width = 0.35\textwidth]{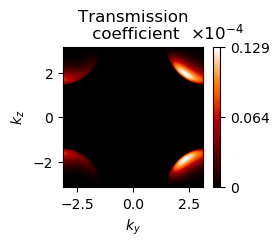}
\caption{Transmission coefficient at $E_F = 5t$, far away from the Weyl points and for $\Delta = 0.1t$ and $M_L = M_R = 1$. The transmission coefficients are orders of magnitude small than those where $E_F$ lies near Weyl points.}
\label{fig:appdx_a}
\end{figure}

Figure \ref{fig:appdx_a} shows TCs at $E_F = 5t$ far away from Weyl points and for $\Delta = 0.1t$ and $M_L = M_R = 1$. The skewness of TCs is weaker than that when $E_F$ is near Weyl points in Fig.\ref{fig:fig2}. Meanwhile, TCs here are 4 order of magnitude smaller than TCs near Weyl points, leading to a very small $g_{yx}$ compared to that in Fig.\ref{fig:fig3}. In contrast, the bulk $\sigma_{xy}$ only reduces by 2 order of magnitude from $ 0.35 e^2/ha_0$ when $E_F = 0$ to $0.0055 e^2/ha_0$ when $E_F = 5t$.
These suggest that the impact of DW on Hall effect is small away from Weyl points and becomes significant when $E_F$ lies near Weyl points.

\section{Computation of conductance}
\label{app:B}
In a bias voltage $\Delta V_x$, the current density can be computed by using the scattering states. The left-incident states are associated with a Fermi distribution function $f_L(E) = f(E-(E_F - e\Delta V_x) )$, while $f_R(E) = f(E - E_F)$ for the right-incident states. This is because the left-incident states are in equilibrium with a reservoir at a different potential energy due to $\Delta V_x$. The current density $j_{\nu}$ for $\nu = x, y$ is given by
\bea
j_{\nu} &=& -e \int \frac{\td \bk_{\parallel} \td E}{(2\pi)^3} [\bra{\Psi_{L; E\bk_{\parallel} \alpha}} \hat{v}_{\nu} \ket{\Psi_{L; E\bk_{\parallel} \alpha}} f_L(E) \nonumber \\
&&+ \bra{\Psi_{R; E \bk_{\parallel} \tilde{\alpha}}} \hat{v}_{\nu} \ket{\Psi_{R; E\bk_{\parallel}\tilde{\alpha}}} f_R(E)],
\label{eq:current}
\eea
where we have identified $\frac{1}{V}\sum_{\bk} = \int \frac{\td \bk}{(2\pi)^3} = \int \frac{\td \bk_{\parallel} \td E}{(2\pi)^3}\frac{1}{|\td E/ \td k_x|}$. The last identification is not strictly rigorous since the denominator $\td E/\td k_x$, i.e. the group velocity in the x-direction, is spatially dependent.  
The proper way is to have a normalization factor in the scattering states \cite{Nazarov2009, Trott1989}, which is done by attaching a prefactor $\frac{1}{\sqrt{\hbar |v_{x\alpha}|}}$ to Eq.(\ref{eq:left_incident}) and $\frac{1}{\sqrt{\hbar |v_{x\tilde{\alpha}}|}}$ to Eq.(\ref{eq:right_incident}). It ensures the anticommutators of the creation and annihilation operators, $\{\Psi_{D; E\bk_{\parallel}\alpha}, \Psi^{\dagger}_{D';E'\bk'_{\parallel}\alpha'}\} = \delta_{DD'}\delta(E-E')\delta(\bk_{\parallel} - \bk'_{\parallel})\delta_{\alpha\alpha'}$, where $D, D' = L, R$ denote the left- or right-incident states. We have included this normalization factor in Eq.(\ref{eq:current}). From these, we obtain
\begin{widetext}
\bea
\label{eq:current_simple}
j_{\nu} &=& -e \int \frac{\td \bk_{\parallel} \td E}{(2\pi)^3} \frac{1}{\hbar |v^x_{\bk_{\parallel} \alpha}|}\left[\frac{1}{2} v^{\nu}_{\bk_{\parallel}, \alpha} + \frac{1}{2} v^{\nu}_{\bk_{\parallel}, \beta} |r^{\alpha \rightarrow \beta}_{L;\bk_{\parallel}}|^2 + \frac{1}{2} v^{\nu}_{\bk_{\parallel}, \tilde{\beta}} |t^{\alpha \rightarrow \tilde{\beta}}_{L;\bk_{\parallel}}|^2 \right](f_L(E) - f_R(E) ),\\
	&=& -\frac{e}{2} \int \frac{\td \bk_{\parallel} \td E}{(2\pi)^3} \frac{1}{\hbar |v^x_{\bk_{\parallel} \alpha}|}\left[v^{\nu}_{\bk_{\parallel}, \alpha} + v^{\nu}_{\bk_{\parallel}, \beta} |r^{\alpha \rightarrow \beta}_{L;\bk_{\parallel}}|^2 + v^{\nu}_{\bk_{\parallel}, \tilde{\beta}} |t^{\alpha \rightarrow \tilde{\beta}}_{L;\bk_{\parallel}}|^2 \right] (-e\Delta V_x\delta(E - E_F)),\\
j_{\nu}	&=& \frac{e^2 \Delta V_x}{2(2\pi)^3\hbar} \int \td \bk_{\parallel} \left[\frac{v^{\nu}_{\bk_{\parallel\alpha}}}{|v^x_{\bk_{\parallel\alpha}}|} + \frac{v^{\nu}_{\bk_{\parallel\beta}}}{|v^x_{\bk_{\parallel\beta}}|} R^{\alpha \rightarrow \beta}_{L;\bk_{\parallel}} + \frac{v^{\nu}_{\bk_{\parallel\tilde{\beta}}}}{|v^x_{\bk_{\parallel\tilde{\beta}}}|} T^{\alpha \rightarrow \tilde{\beta}}_{L;\bk_{\parallel}} \right]_{E = E_F},
\eea
\end{widetext}
where the sums over $\alpha, \tilde{\alpha}, \beta, \tilde{\beta}$ are implicit. We have used the condition that the current density is zero at zero bias voltage when $f_L = f_R$ to arrive at Eq.\ref{eq:current_simple}. The factors $1/2$ in the square brackets originate from the expectation value of the velocity operator evaluated from half of the space.
The difference $f_L(E) - f_R(E) \approx -e\Delta V_x\delta(E - E_F)$ is obtained at the zero temperature limit, and $E_F$ is the Fermi level. The longitudinal current density $j_x$ is given by
\bea
j_x &=& \frac{e^2 \Delta V_x}{2(2\pi)^3\hbar} \int \td \bk_{\parallel} \left[1 - R^{\alpha \rightarrow \beta}_{L;\bk_{\parallel}} + T^{\alpha \rightarrow \tilde{\beta}}_{L;\bk_{\parallel}} \right]_{E = E_F}, \nonumber\\
j_x &=& \frac{e^2 \Delta V_x}{(2\pi)^3\hbar} \int \td \bk_{\parallel} \left[ T^{\alpha \rightarrow \tilde{\beta}}_{L;\bk_{\parallel}} \right]_{E = E_F}.
\eea
This is the familiar Landauer formula which relates conductance to transmission coefficients.

\bibliography{refs}

\end{document}